# Modeling of Anisotropic Two-Dimensional Materials Monolayer HfS$_2$ and Phosphorene MOSFETs

Jiwon Chang

SEMATECH, 257 Fuller Rd #2200, Albany, NY 12203, USA

Ballistic transport characteristics of metal-oxide semiconductor field effect transistors (MOSFETs) based on anisotropic two-dimensional (2-D) materials monolayer HfS$_2$ and phosphorene are explored through quantum transport simulations. We focus on the effects of the channel crystal orientation and the channel length scaling on device performances. Especially, the role of degenerate conduction band (CB) valleys in monolayer HfS$_2$ is comprehensively analyzed. Benchmarking monolayer HfS$_2$ with phosphorene MOSFETs, we predict that the effect of channel orientation on device performances is much weaker in monolayer HfS$_2$ than in phosphorene due to the degenerate CB valleys of monolayer HfS$_2$. Our simulations also reveal that, at 10 nm channel length scale, phosphorene MOSFETs outperform monolayer HfS$_2$ MOSFETs in terms of the on-state current. However, it is observed that monolayer HfS$_2$ MOSFETs may offer comparable, but a little bit degraded, device performances as compared with phosphorene MOSFETs at 5 nm channel length.

# I. INTRODUCTION

Recent years have witnessed an upsurge of interest in two-dimensional (2-D) layered materials, particularly transition metal dichalcogenides (TMDs). Since the thickness of TMDs can be pushed down to less than a few nanometers, TMDs are promising alternative channel materials for realizing ultra-thin body metal-oxide-semiconductor field effect transistors (MOSFETs) which are robust to short channel effects. First experimental demonstration of monolayer $MoS_2$ MOSFETs suggested the possibility of monolayer TMDs for the electronic device application [1] and ignited extensive following researches on Mo- and W-based TMDs such as $MoS_2$, $MoSe_2$, $MoTe_2$, $WS_2$ and $WSe_2$ [2,3,4,5]. Continuous efforts to seek for other 2-D materials lead to the discovery of anisotropic 2-D material phosphorene [6]. Unlike Mo- and W-based TMDs, phosphorene is characterized with its highly anisotropic band structures [6,7,8]. Theoretical investigations of phosphorene MOSFETs have been performed [9,10] and reported that the unique anisotropy of phosphorene band structures is advantageous to improve ballistic device performances over the nearly isotropic 2-D material monolayer $MoS_2$ MOSFETs. Very recently, anisotropic 2-D materials other than phosphorene, Hf- and Zr-based TMDs, have been explored by density functional theory (DFT) calculations [11,12]. Monolayers of Hf- and Zr-based TMDs such as $HfS_2$, $HfSe_2$, $ZrS_2$ and $ZrSe_2$ are quite different with those of Mo- and W- based TMDs because of different crystal symmetry and atomic bonding [11,12]. They have indirect band gaps with three-fold degenerate conduction band (CB) valleys whose dispersions around the minimum of CB are highly anisotropic. Among Hf- and Zr-based TMDs, monolayer $HfS_2$ and monolayer $ZrS_2$ exhibit sizable band gaps, larger than 1 eV, suitable for MOSFETs applications [11,12]. In this work, we presents a comprehensive computational study of monolayer $HfS_2$ MOSFETs through the ballistic quantum transport simulations. We discuss the influence of band structure anisotropy of monolayer HfS$_2$ on device performances as well as the scaling behavior of monolayer HfS$_2$ MOSFETs. Simulations of phosphorene MOSFETs with the same device geometry and biasing conditions are also carried out to benchmark key performance metrics with monolayer HfS$_2$ MOSFETs.

## II.  COMPUTATIONAL APPROACH

Figure 1(a) shows the top view of monolayer HfS$_2$ atomic structure and the corresponding 1$^{st}$ Brillouin Zone (BZ). Similar to MoS$_2$, HfS$_2$ is a layered material composed of vertically stacked S-Hf-S layers through van der Waals forces. Each single S-Hf-S layer consists of two hexagonal planes of S atoms and an intermediated hexagonal plane of Hf atoms interacting through ionic-covalent interactions. Difference between MoS$_2$ and HfS$_2$ is that hexagonal planes of S are A-A stacked in MoS$_2$ while A-B stacked in HfS$_2$. Electronic structure calculations and structure optimization of monolayer HfS$_2$ were performed through DFT calculations by OPENMX [13] using the linear combination of pseudoatomic orbital method with the local density approximation (LDA) to describe the exchange-correlation [14]. We constructed monolayer HfS$_2$ structure by using the in-plain lattice parameter ($a$ = 3.622 Å) from experiments [15] since it was reported that 2-D materials remain very close to their 3-D parents [16]. We fixed the in-plane lattice value and relaxed atoms with a force tolerance of 0.001 Hartree/Bohr. The cutoff energy of 300 Ryd and 7×7×1 $k$-mesh for the BZ integrations were used for the structure optimization as well as band structure calculations. Similar to the other plane wave based DFT calculation [11,12], we found the highly anisotropic CB with its minimum located at the M-point as in Figure 1(d). The calculated indirect band gap is 1.13 eV in close agreement with the previous theoretical prediction [11]. Experimental verification of band gap has not been demonstrated yet. However, a band gap size has limited effects on the MOSFETs simulation since both CB and valence band (VB) are considered simultaneously only in the calculation of gate-induced drain leakage (GIDL) current which has not been observed in the simulated range of $V_{GS}$ in this work. There are three CB valleys in the 1$^{st}$ BZ of monolayer HfS$_2$ as seen in Figure 1(a). We extracted electron effective masses of three valleys in two directions $M$ ($\Gamma \rightarrow M$) and $K$ ($\Gamma \rightarrow K$) by parabolic fitting of the band structure and listed them in Figure 1(b). In the direction $M$, valley 1 has a very heavy effective mass $m_e^* = 3.05 \times m_e$ with a light transverse effective mass leading to a low density of states (DOS) while the other two valleys have light ones $m_e^* = 0.35 \times m_e$. In the other direction $K$, valley 2 exhibits a light effective mass $m_e^* = 0.25 \times m_e$ with a heavy transverse effective mass for a high DOS and the other two valleys show heavier effective masses $m_e^* = 0.81 \times m_e$.

In the simulated device structure shown in Figure 1(c), monolayer HfS$_2$ with a dielectric constant $\kappa = 6.2$ [17] was used as a channel material with the 3 nm HfO$_2$ ($\kappa = 25$) gate oxide and the 10 nm SiO$_2$ ($\kappa = 3.9$) substrate. Semi-infinite source and drain were n-type doped with a doping concentration of $2 \times 10^{13}$/cm$^2$. We varied the channel length $L_{Ch}$ from 10 to 5 nm to examine scaling behavior. To describe electronic transport through monolayer HfS$_2$, we performed self-consistent ballistic quantum transport simulations using tight-binding (TB) hopping potentials. TB potentials were obtained through maximally localized Wannier functions (MLWFs) calculated directly from the DFT Kohn-Sham orbitals and potential using OPENMX [18]. As in Figure 1(d), TB potentials accurately reproduce original DFT band structures. We used the scattering matrix approach to propagate injected eigenmodes from semi-infinite source and drain through the device [19]. Transport equations were solved iteratively together with Poisson's equation for the self-consistency between the charge density and the electrostatic potential. The total current was calculated within the Landauer–Büttiker formalism. Two transport directions $M$ ($\Gamma \rightarrow M$) and $K$ (Γ→K) of monolayer HfS$_2$ in Figure 1(a) were investigated in the quantum transport simulations. We benchmarked performances of monolayer HfS$_2$ MOSFETs with those of phosphorene MOSFETs in the same device structure and biasing conditions. TB Hamiltonian of phosphorene used in our previous modeling work [20] was adopted for quantum transport calculations. We considered two transport directions $X$ (Γ→X) and $Y$ (Γ→Y) for light and heavy effective masses $m_e^* = 0.115 \times m_e$ and $m_e^* = 1.17 \times m_e$ in phosphorene, respectively, in Figure 1(e).

## III. RESULTS AND DISCUSSION

We analyze simulation results in terms of effective masses in both transport and transvers directions. Even if the effective mass becomes irrelevant to describe the ballistic conductance between two terminals within the Landauer–Büttiker formalism, understanding MOSFETs performances with the effective mass is still valid as the current is controlled by the gate via self-consistent electrostatic field. Simulation results of 10, 7 and 5 nm channel length monolayer HfS$_2$ and phosphorene MOSFETs are presented in Figure 2 and 3. Figure 2(a) compares transfer characteristics of monolayer HfS$_2$ and phosphorene MOSFETs with 10 nm channel length in different transport directions at $V_{DS} = 0.5$ V. We adjusted $V_{GS}$ such that the off-state ($V_{DS} = 0.5$ V and $V_{GS} = 0$ V) current $I_{OFF}$ is 100 nA/μm according to the ITRS requirement for high performance logic devices [21]. For both monolayer HfS$_2$ and phosphorene MOSFETs, irrespective of transport directions, good subthreshold behavior and limited short-channel effects are observed in 10 nm channel length device. Subthreshold slope (*SS*) and drain-induced barrier lowering (DIBL) are estimated to ~ 70 mV/dec and ~ 50 mV/V, respectively in Figure 3(a) and 3(b). As $V_{GS}$ increases, transfer characteristics start to show dependency on the channel material and the transport direction. At the on-state ($V_{DS} = V_{GS} = 0.5$ V), phosphorene MOSFETs offers the largest on-state current $I_{ON}$ ~ 3000 µA/µm in X-direction due to the light transport effective mass and the heavy transverse effective mass while the smallest $I_{ON}$ ~ 1200 µA/µm in Y-direction because of the heavy transport and light transverse effective masses as similar to the previous reports [9,10]. In monolayer HfS$_2$ MOSFETs, both transport directions M and K result in the quite similar level of $I_{ON}$. $I_{ON}$ in M-direction is slightly higher (~ 1820 µA/µm) than in K-direction (~ 1730 µA/µm). Compared with phosphorene MOSFETs, a very narrow range of $I_{ON}$ (1730 ~ 1820 µA/µm) depending on the channel crystal orientation is obtainable in monolayer HfS$_2$ while a much wider range (1200 ~ 3000 µA/µm) in phosphorene. $I_{ON}$ in monolayer HfS$_2$ reaches roughly 57 ~ 60 % of the maximum $I_{ON}$ in phosphorene MOSFETs. Therefore, as shown from the plot of $I_{ON}$ as a function of $I_{ON}/I_{OFF}$ ratio in Figure 3(c), we can achieve a larger $I_{ON}$ in phosphorene than in monolayer HfS$_2$ with the same $I_{ON}/I_{OFF}$ ratio and similar subthreshold characteristics ((Figure 3(a) and 3(b)) by properly adjusting the channel crystal orientation.

    For deeper understanding of the current transport, we plot CB edge profiles and corresponding energy resolved current densities for 10 nm channel length MOSFETs at the on-state in Figure 4. In the plots, the source and drain Fermi levels are indicated by $E_{FS}$ and $E_{FD}$, respectively. For the given source and drain doping concentration of $2\times10^{13}$/cm$^2$, positions of $E_{FS}$ and $E_{FD}$ relative to CB edge are higher in phosphorene than in monolayer HfS$_2$ since monolayer HfS$_2$ has three degenerate valleys with the heavier effective masses compared to only one valley with the relatively lighter effective masses in phosphorene. From Figure 4(a) and 4(b), phosphorene MOSFETs in X-direction provide improved $I_{ON}$ because the light effective mass enhances the injection velocity and the heavy transverse mass leads to the high DOS as discussed above [9,10]. For monolayer HfS$_2$, the current density originating from each valley in Figure 1(a) is shown separately in Figure 4(c) and 4(d). We consider current up to 500 meV above $E_{FS}$ because

current becomes negligibly small and it becomes difficult to discriminate one valley to the others above that energy level. From Figure 4(c), current from valley 1 is relatively small compared to those from the other valleys due to its heavy effective mass $m_e^* = 3.05 \times m_e$ (Figure 1(b)). The contribution of each valley current to $I_{ON}$ is more clearly seen in Figure 5. Only about 13.7% of $I_{ON}$ comes from valley 1 and the rest originates from valley 2 and 3. As we change the transport direction from *M* to *K*, the transport effective mass for valley 1 is significantly reduced, but still heavy ($m_e^* = 0.81 \times m_e$), while that for valley 3 increases and eventually becomes same with the value of valley 1 as in Figure 1(b). For valley 2 in *K*-direction, on the other hand, it is aligned to have the lightest transport and the heaviest transverse effective masses in monolayer $HfS_2$ CB as explained in Figure 1(a). Therefore, valley 2 contributes the most among three valleys to $I_{ON}$ in the current density plots of Figure 4(d). As observed in Figure 5, the valley 2 current accounts for the biggest portion of $I_{ON}$ (more than 58 %) while valley 1 and 3 provide about 21 % of $I_{ON}$ for each. From comparing components of $I_{ON}$ between transport directions *M* and *K*, the valley 3 current in *M*-direction is lowered by more than half in *K*-direction owing to increasing transport and decreasing transverse effective masses. On the other hand, the valley 1 and 2 current boost more than 50 % and 36 %, respectively, in *K*-direction because of lighter transport and heavier transverse effective masses. As a result, even though the current from valley 3 decreases, the amount of current increase from the other valleys, particularly valley 2, compensates the decrease, thereby resulting in the slight increase of overall $I_{ON}$ in the transport direction *K*. Unlike phosphorene, even if monolayer $HfS_2$ has anisotropic CB, it does not exhibit high dependency of device performances on the orientation of channel material in 10 nm channel length scale due to its degenerate CB valleys.

To examine the scaling limit of monolayer $HfS_2$ and phosphorene MOSFETs, we simulated 7 and 5 nm channel length devices with the same other device parameters under the same biasing conditions. Figure 3(a) and 3(b) show the scaling behavior of *SS* and DIBL for monolayer $HfS_2$ and phosphorene MOSFETs. As the channel length scales down, subthreshold characteristics are substantially degraded due to the short channel effects as well as the source-to-drain direct tunneling. Below 10 nm channel length scale, the choice of channel material and the transport direction affects subthreshold characteristics as summarized in Figure 3(a) and 3(b). In phosphorene, at 5 nm, *SS* and DIBL increase up to 125 mV/dec and 206 mV/V in *X*-direction while up to 82 mV/dec and 127 mV/V in *Y*-direction. Degradation in the transport direction *Y* is suppressed due to the heavy effective mass which effectively blocks the source-to-drain direct tunneling as discussed in the previous work [10]. Monolayer $HfS_2$ MOSFETs also exhibits significant degradations of *SS* and DIBL. At 5nm channel length scale, *SS* of 87 ~ 94 mV/dec and DIBL of 165 ~ 172 mV/V are observed, depending on the transport direction. Degradation is slightly more severe in the transport direction *K* than in *M*. In comparison with phosphorene MOSFETs, the transport directional dependency of subthreshold characteristics is not substantial in monolayer $HfS_2$ MOSFETs because of degenerate CB valleys as indicated from Figure 3(b) and 3(c). DIBL and *SS* of monolayer $HfS_2$ MOSFETs lie somewhere between the maximum and minimum values of phosphorene MOSFETs. *SS* is managed better in monolayer $HfS_2$ than in phosphorene aligned in *X*-direction, suggesting the reduced source-to-direct tunneling. Transfer characteristics of 5 nm channel length devices at $V_{DS} = 0.5$ V are plotted in Figure 2(b). Here, $V_{GS}$ is not adjusted to yield $I_{OFF}$ of 100 nA/μm at $V_{GS} = 0$ V. Instead, we used the same $V_{GS}$ range applied in 10 nm channel length device to check the threshold voltage $V_T$ roll-off. As discussed above, the highest subthreshold current and the most severe $V_T$ roll-off are exhibited for phosphorene in the transport direction *X* because of the largest *SS* in Figure 3(a). Similar to 10 nm channel length device, a wide range of $I_{ON}$ is achievable in phosphorene via tuning the transport direction while changing transport direction does not have a critical influence on $I_{ON}$ in monolayer HfS$_2$. Compared with 10 nm device, $I_{ON}$ improves by nearly 60 % and 35 % in *X*- and *Y*-directions, respectively, in phosphorene and reaches ~ 4800 µA/µm and ~ 1615 µA/µm. In monolayer HfS$_2$, roughly 40 % more $I_{ON}$ is obtained for both *M*- and *K*-directions at the on-state. This enhancement of $I_{ON}$ in 5nm channel length device is attributed not only to the electrostatic short channel effects but also to the source-to-drain direct tunneling. Especially for phosphorene MOSFETs in *X*-direction, the source-to-drain tunneling current becomes considerable because of the light transport effective mass, thereby leading to the biggest increase of $I_{ON}$ [10]. Even though monolayer HfS$_2$ has at least one valley with a light effective mass in both *M*- and *K*-directions (Figure 1(b)), valley with the light effective mass does not result in the significant degradation of device performances in overall as for phosphorene in *X*-direction.

To further evaluate subthreshold characteristics, CB edge profiles and corresponding energy resolved current densities for 5 nm channel length devices at the off-state ($V_{GS} = 0$ V) and at $V_{GS} = -0.2$ V are shown in Figure 6. Figure 6(a) and 6(b) explicitly presents the difference of source-to-drain tunneling current between *X*- and *Y*- directions of phosphorene MOSFETs at the off-state. Phosphorene directed in *Y*-direction suppresses the tunneling current more efficiently than in *X*-direction, offering the best subthreshold device performances in Figure 2(b) and Figure 3(a) and 3(b). $I_{OFF}$ of monolayer HfS$_2$ is investigated through calculating the current density from each valley in Figure 6(c) and 6(d). Figure 7(a) shows the portions of $I_{OFF}$ for source-to-drain tunneling (TN) and thermionic emission (TE) currents as well as the contribution of each valley to TN and TE currents. From Figure 7(a), TN (grey solid) and TE (grey striped) currents supply about 60 % and 40 % of $I_{OFF}$ in *M*-direction, respectively. Valley 1 (red) contributes the smallest amount of both TN and TE currents among three valleys because its heavy effective mass $m_e^* = 3.05 \times m_e$ in Figure 1(b) lowers the tunneling efficiency and the injection velocity as well. If we change the transport direction from *M* to *K*, TN current increases by roughly 20 % while only about 8 % boost in TE current as seen in Figure 7(a). As a result, overall 15 % more $I_{OFF}$ is observed in *K*-direction. As discussed in 10 nm device, aligning monolayer $HfS_2$ in *K*-direction puts valley 2 in the optimal orientation for the maximum current (the lightest transport and heaviest transverse effective masses). Therefore, the major contributor to $I_{OFF}$ in *K*-direction is from valley 2 (green) whose components of TN and TE currents are as high as 87 % and 55 %, respectively in Figure 7(a). Figure 6(d) also confirms that the valley 2 current is dominant, especially for TN current below the top of the potential barrier. The valley 1 current (red) also increases, but remains relatively small because the transport effective mass is still heavy ($m_e^* = 0.81 \times m_e$) in *K*-direction. On the other hand, valley 3 (blue) provides less current in *K*-direction due to the heavier transport effective mass than in *M*-direction. Essentially the same as the discussion of the on-state current in 10 nm channel length $HfS_2$ MOSFETs, the three-fold valley degeneracy of CB in monolayer $HfS_2$ diminishes the impact of channel orientation on device performances in the subthreshold regime. However, at 5 nm scale where the quantum mechanical source-to-drain tunneling occurs, changing transport direction starts to make a slight difference. From Figure 7(a), $I_{OFF}$ is 15% more in the transport direction *K* mainly due to the increase of TN current from valley 2. This distinction between *M*- and *K*-directions becomes clearer at lower $V_{GS}$ where TN current takes up the subthreshold current more dominantly. Figure 6(e) and 6(f) are CB edges and current densities of monolayer $HfS_2$ MOSFETs for *M*- and *K*-directions, respectively, at $V_{GS} = -0.2$ V. TN currents from valley 1 for the transport direction *M* and valley 1 and 3 for the transport direction *K* are well suppressed because of the heavy transport effective masses. On the other hand, valley 2 and 3 constitute most TN current in Figure 6(e) while only valley 1 supplies most of it as in Figure 6(f).

However, the amount of TN current from valley1 in *K*-direction exceeds the sum of current from valley 2 and 3 in *M*-direction. Components of current at $V_{GS} = -0.2$ V is further analyzed in Figure 7(b), indicating a bigger relative difference of current between *M*- and *K*-directions than at the off-state ($V_{GS} = 0$ V) in Figure 7(a). Comparing Figure 7(a) with 7(b) reveals that changing transport direction from *M* to *K* enhances TN current by about 82 % at $V_{GS} = -0.2$ V compared with 20 % at $V_{GS} = 0$ V, and TE current by about 20 % at $V_{GS} = -0.2$ V compared with 8 % at $V_{GS} = 0$ V. This significant increase of TN current in *K*-direction at $V_{GS} = -0.2$ V is mainly attributed to the huge boost of TN current from valley 2 as shown in Figure 7(b). Monolayer $HfS_2$ directed in *K*-direction becomes more vulnerable to the source-to-drain direct tunneling since valley 2 aligned for the lightest transport and the heaviest transverse masses maximizes the tunneling efficiency. Therefore, subthreshold device performances of monolayer $HfS_2$ are further degraded when monolayer $HfS_2$ is oriented to *K*-direction, hence creating a difference in the subthreshold current between two transport directions. We can expect an even stronger dependency of device performances of monolayer $HfS_2$ MOSFETs on the transport direction below 5 nm channel length scale where the source-to-drain direct tunneling has more sizable impact.

## IV. CONCLUSION

We examined device performances of MOSFETs based on the anisotropic 2-D material monolayer $HfS_2$ through the ballistic quantum transport simulations. The dependency of device performances on the transport direction and the scaling behavior were assessed and benchmarked with phosphorene MOSFETs. At 10 nm channel length scale, both monolayer $HfS_2$ and phosphorene MOSFETs offer excellent subthreshold characteristics regardless of the transport direction. At the on-state, however, phosphorene can provide much higher level of $I_{ON}$ than monolayer HfS$_2$ when the light effective mass direction is used for the transport direction in phosphorene MOSFETs. Even though both monolayer HfS$_2$ and phosphorene have anisotropic band structures, improving device performances through tuning the channel crystal orientation is expected only in phosphorene because three-fold degenerate CB valleys in monolayer HfS$_2$ reduce the effect of channel orientation. As the channel length is scaled down, the most substantial degradation of subthreshold characteristics is observed in phosphorene MOSFETs aligned in the light transport effective mass direction since it suffers from the severe source-to-drain direct tunneling. On the other hand, directing phosphorene to the heavy effective direction yields the best subthreshold characteristics and comparable, but slightly more degraded, subthreshold characteristics are achievable with monolayer HfS$_2$. Below 5 nm channel length, in the presence of source-to-drain direct tunneling, monolayer HfS$_2$ starts to exhibit the channel orientation dependency in the subthreshold regime because the source-to-drain tunneling drastically increases or decreases, depending on the transport effective mass of each valley in monolayer HfS$_2$.

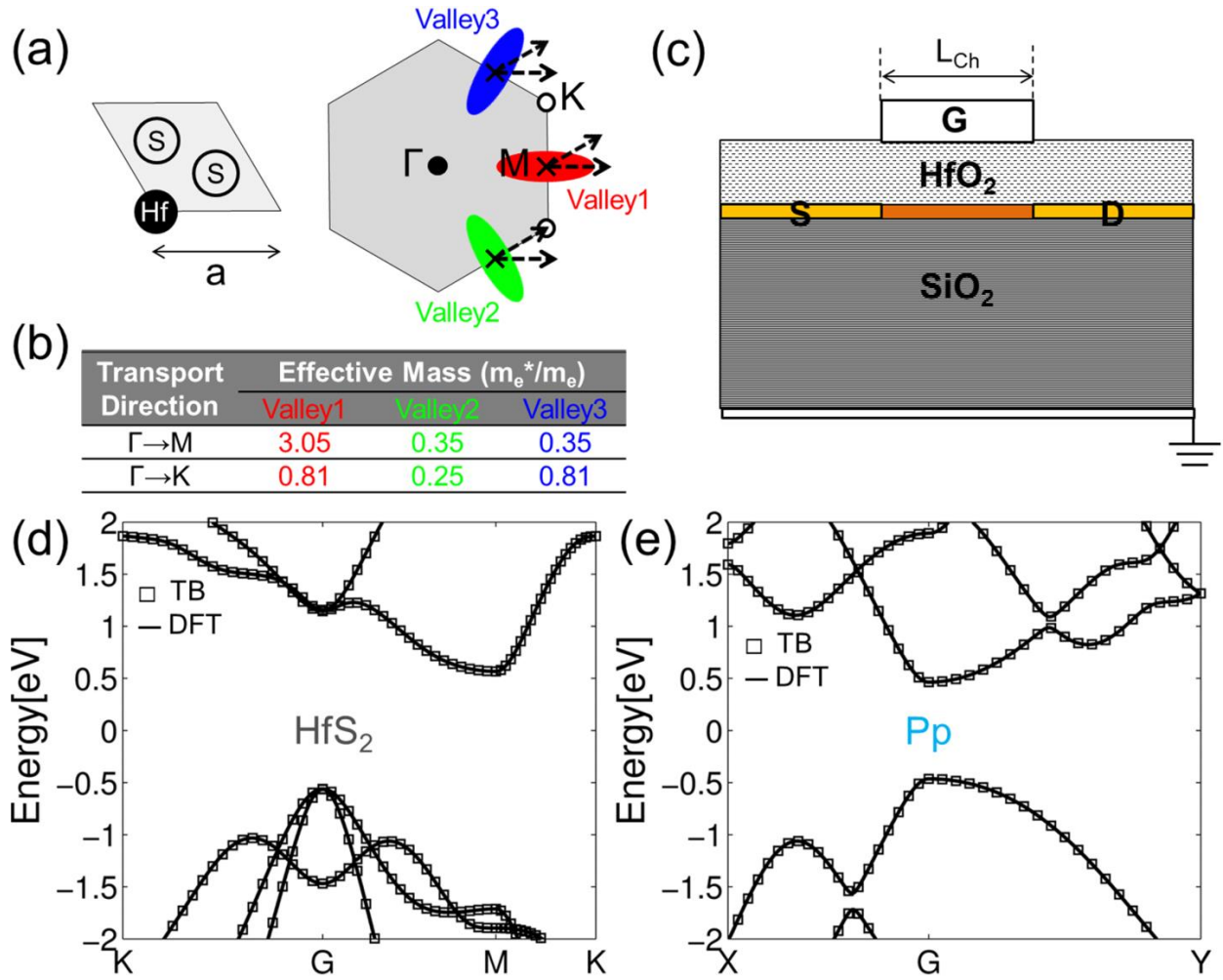

**FIG. 1**. (a) Top view of monolayer HfS$_2$ showing a primitive hexagonal unit cell and corresponding 1$^{st}$ BZ with high symmetric points. Three lowest conduction band valleys are located at three M points. (b) Electron effective masses along two transport directions for each valley in (a). (c) Schematic of simulated device structure. The nominal device parameters are as follows: HfO$_2$ ($\kappa$ = 25) gate oxide thickness = 3 nm, channel length L$_{Ch}$ = 5, 7 and 10 nm, n-type doping density of source and drain = 2×10$^{13}$ cm$^{-2}$, and SiO$_2$ oxide thickness = 10 nm. Band structures of (d) monolayer HfS$_2$ and (e) phosphorene from DFT (solid lines) and from TB Hamiltonian (squares) along the high symmetric points in the hexagonal BZ and rectangular BZ, respectively.

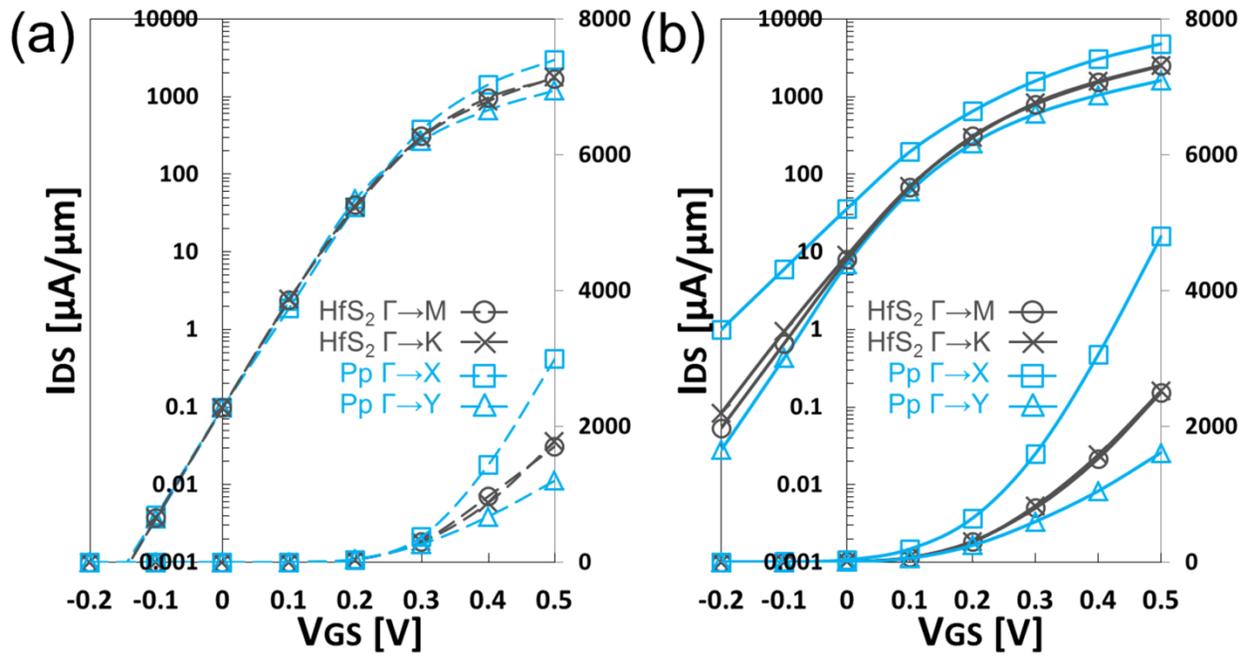

**FIG. 2**. Transfer characteristics of (a) 10 nm and (b) 5 nm channel length monolayer HfS$_2$ and phosphorene MOSFETs in different transport directions at $V_{GS}$ = 0.5 V.

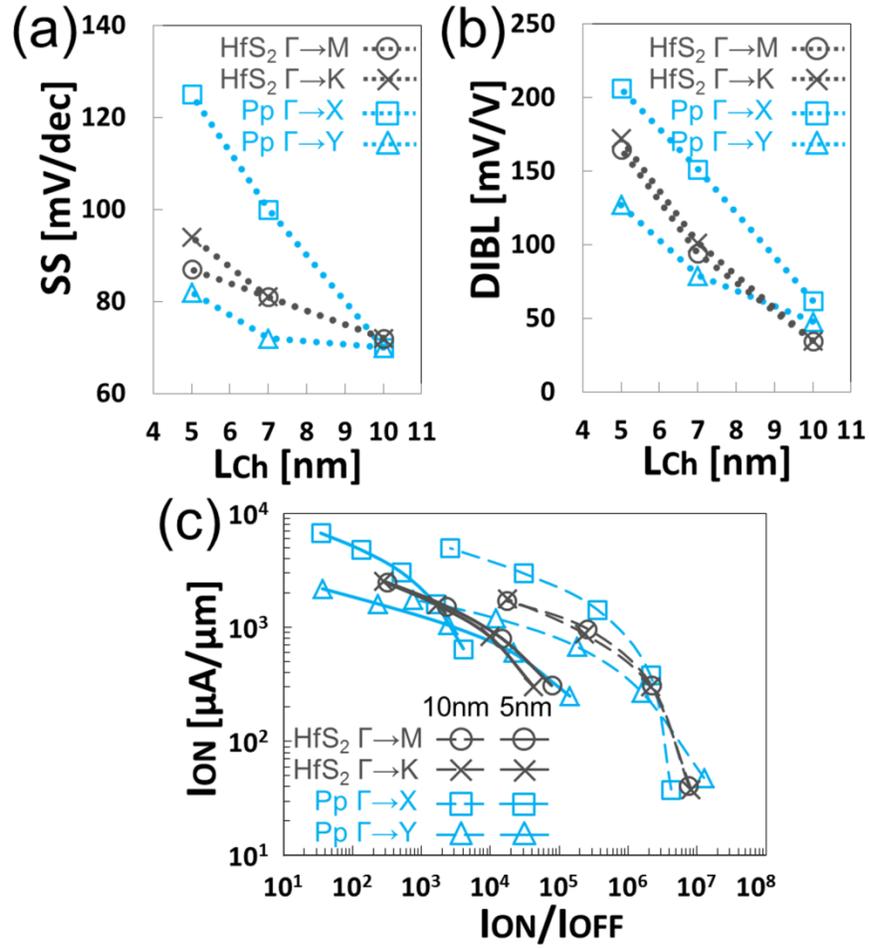

**FIG. 3**. (a) *SS*, (b) DIBL and (c) $I_{ON}$ vs. $I_{ON}/I_{OFF}$ ratio of monolayer $HfS_2$ and phosphorene MOSFETs for different channel lengths in different transport directions.

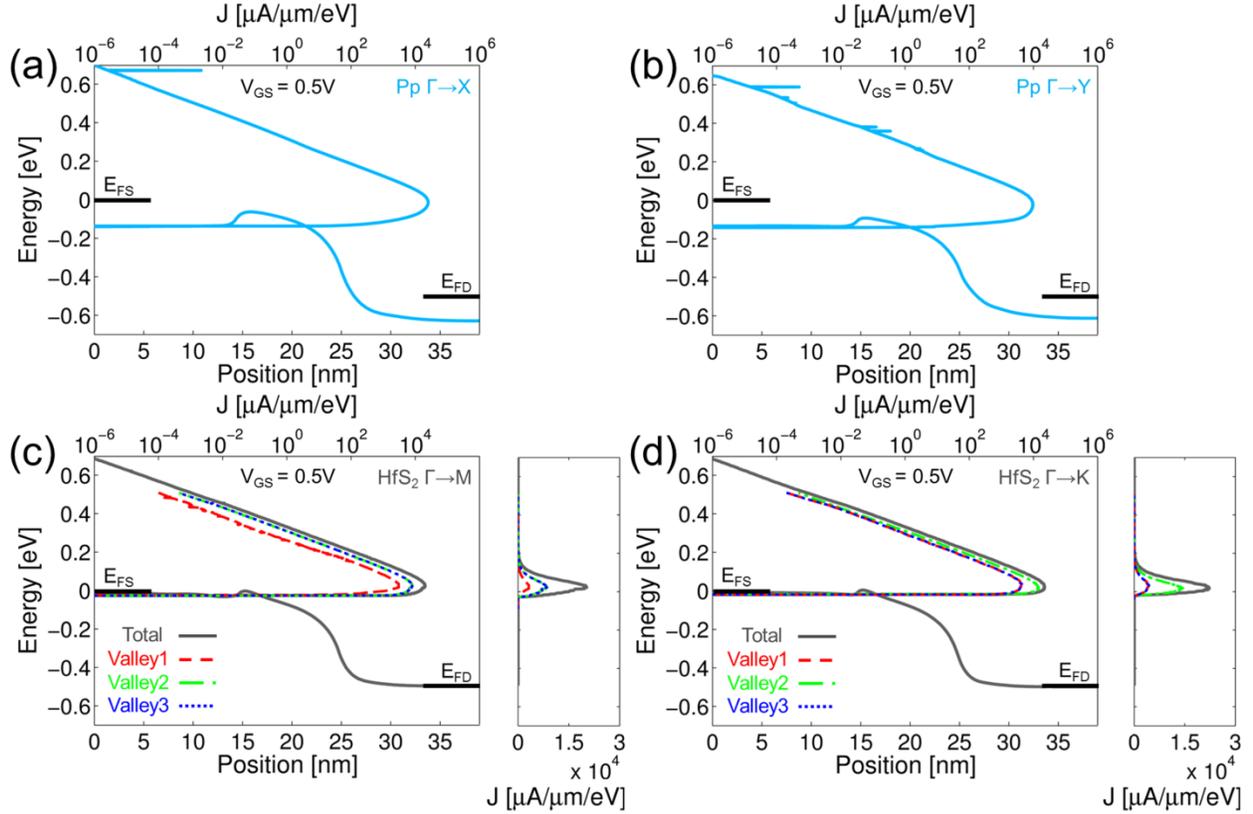

**FIG. 4**. CB edge profiles and corresponding energy resolved current densities for 10 nm channel length phosphorene MOSFETs in transport directions (a) Γ→X and (b) Γ→Y and for 10 nm channel length monolayer HfS$_2$ MOSFETs in transport directions (c) Γ→M and (d) Γ→X at $V_{GS}$ = 0.5V. $E_{FS}$ and $E_{FD}$ represent Fermi levels in the source and drain, respectively. Current density up to 500 meV above the source Fermi level from each valley of monolayer HfS$_2$ in Figure 1(a) is plotted in log (left) and linear (right) scales in (c) and (d).

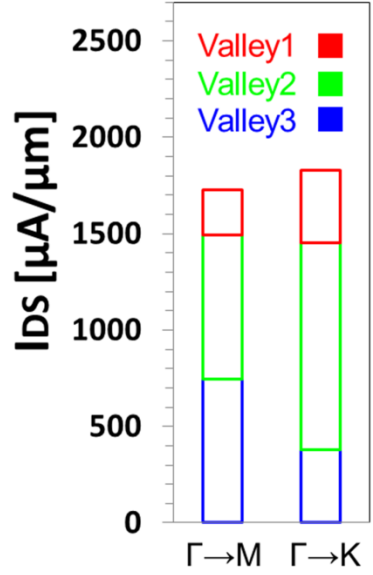

**FIG. 5**. Components of current in 10nm channel length monolayer HfS$_2$ MOSFETs at $V_{GS} = 0.5$V for different transport directions. Red, green and blue bars represent the portions of total current for three valleys in Figure 1(a), respectively.

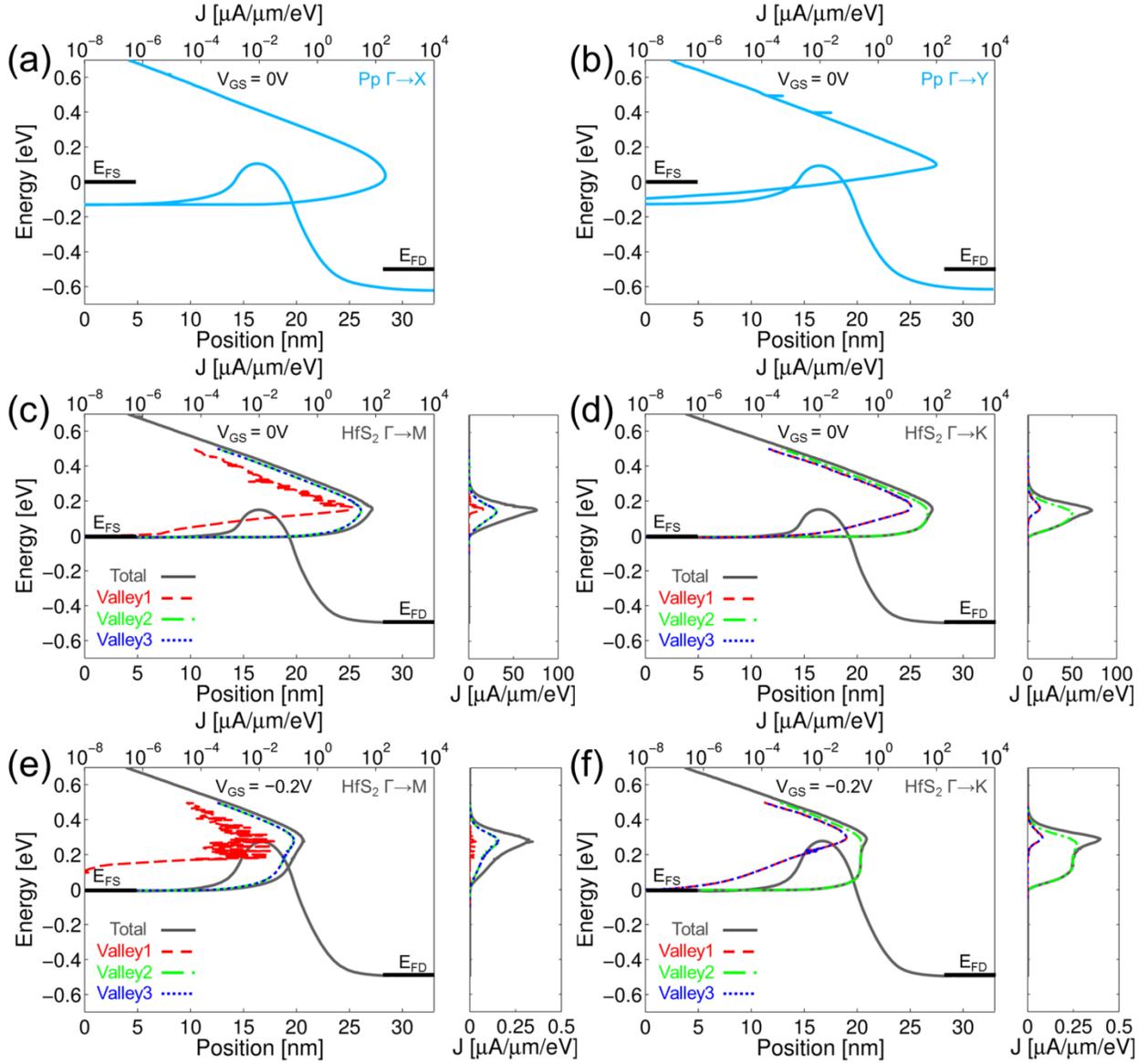

**FIG. 6**. CB edge profiles and corresponding energy resolved current densities for 5nm channel length phosphorene MOSFETs at $V_{GS}$ = 0V in transport directions (a) Γ→X and (b) Γ→Y. CB edge profiles and corresponding energy resolved current densities for 5nm channel length monolayer HfS$_2$ MOSFETs at $V_{GS}$ = 0V in transport directions (c) Γ→M and (d) Γ→X and at $V_{GS}$ = −0.2V in transport directions (e) Γ→M and (f) Γ→X. $E_{FS}$ and $E_{FD}$ represent Fermi levels in the source and drain, respectively. Current density up to 500 meV above the source Fermi

Level from each valley of monolayer HfS$_2$ in Figure 1(a) is plotted in log (left) and linear (right) scales in (c), (d), (e) and (f).

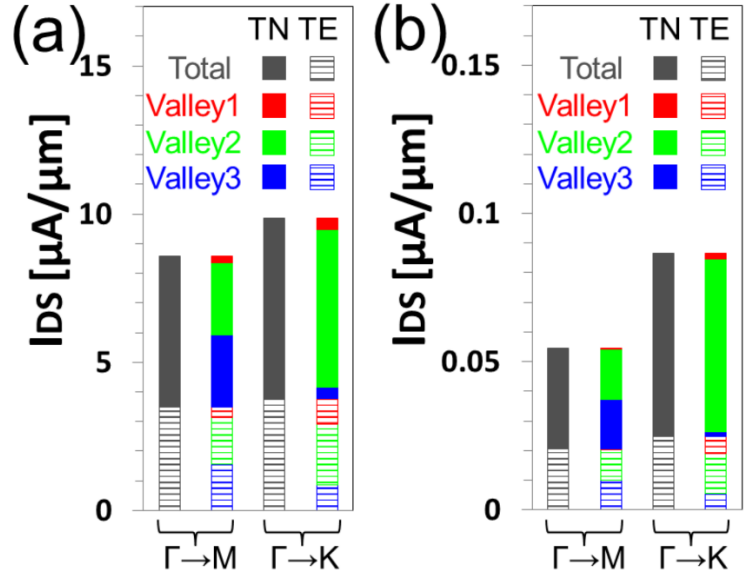

**FIG. 7**. Components of current in 5nm channel length monolayer HfS$_2$ MOSFETs at $V_{GS}$ = (a) 0.0V and (b) −0.2V for different transport directions. Grey solid and striped bars correspond to the portions of total current for source to drain tunneling (TN) and thermionic emission (TE) currents, respectively. Red, green and blue bars represent the portions of TN current or TE current for each valley in Figure 1(a).